\documentclass[conference, 10pt, 1in]{IEEEtran}
\usepackage[font=small,labelfont=bf]{caption}
\usepackage{graphicx, balance}
\usepackage{caption,subcaption}
\usepackage{linguex}
\usepackage[linguistics,edges]{forest}
\usepackage{enumitem}
\usepackage{booktabs}
\usepackage{multirow, makecell}
\usepackage[cmex10]{amsmath}
\usepackage{algorithmic}
\usepackage{url}
\usepackage{csquotes} 
\usepackage{arydshln} 
\usepackage{color, xcolor, soul}
\usepackage{amssymb}
\usepackage{makecell}
\usepackage{cite}
\usepackage{etoolbox}
\usepackage{threeparttable}
\usepackage{hyperref}
\usepackage{pifont}
\usepackage{tikz}
\usetikzlibrary{matrix}

\usepackage{array}
\usepackage{booktabs}
\usepackage{multirow}
\usepackage{tabularx}

\usepackage{rotating}

\newcolumntype{C}[1]{>{\centering\arraybackslash}p{#1}}

\definecolor{StartStopColor}{HTML}{FC427B} 
\definecolor{InputOutputColor}{HTML}{00FFFF} 
\definecolor{ProcessColor}{HTML}{50C878} 
\definecolor{DecisionColor}{HTML}{A7257D}
\definecolor{textcolor}{HTML}{000000}
\definecolor{processtextcolor}{HTML}{000000}

\usepackage{tikz}
\usetikzlibrary{shapes.geometric, arrows}
\tikzstyle{startstop} = [rectangle, rounded corners, minimum width=3cm, minimum height=1cm, text centered, draw=black, text=textcolor, fill=StartStopColor!80]
\tikzstyle{io} = [trapezium, trapezium left angle=70, trapezium right angle=110, text width=5cm, minimum height=1cm, text centered, draw=black, fill=InputOutputColor!50, text=textcolor]
\tikzstyle{process} = [rectangle, text width=8cm, minimum height=1cm, text centered, draw=black, fill=ProcessColor!80, text=textcolor]
\tikzstyle{decision} = [diamond, minimum width=3cm, minimum height=1cm, text centered, draw=black, fill=DecisionColor]
\tikzstyle{arrow} = [thick,->,>=stealth]

\tikzstyle{io_1} = [trapezium, trapezium left angle=70, trapezium right angle=110, text width=6cm, minimum height=1cm, text centered, draw=black, fill=InputOutputColor, text=textcolor]
\tikzstyle{process_1} = [rectangle, text width=5cm, minimum height=1cm, text centered, draw=black, fill=ProcessColor, text=textcolor]

\title{Bridging the Gap in Phishing Detection: \\ A Comprehensive Phishing Dataset Collector}

\author{
    \IEEEauthorblockN{
        Aditya Kulkarni\textsuperscript{*},
        Shahil Manishbhai Patel\textsuperscript{*},
        Shivam Pradip Tirmare\textsuperscript{*},
        Vivek Balachandran\textsuperscript{\#}, and 
        Tamal Das\textsuperscript{*}
    }

    \IEEEauthorblockA{
        \textsuperscript{*}\emph{Indian Institute of Technology, Dharwad, India} \\
        \textsuperscript{\#}\emph{Singapore Institute of Technology, Singapore} \\
        \textsuperscript{*}\{aditya.kulkarni, shahilpatel, shivamtirmare, tamal\}@iitdh.ac.in, \textsuperscript{\#}vivek.b@singaporetech.edu.sg}
}

\begin{document}

\maketitle

\begin{abstract}
   To combat phishing attacks -- aimed at luring web users to divulge their sensitive information -- various phishing detection approaches have been proposed. As attackers focus on devising new tactics to bypass existing detection solutions, researchers have adapted by integrating machine learning and deep learning into phishing detection. Phishing dataset collection is vital to developing effective phishing detection approaches, which highly depend on the diversity of the gathered datasets. The lack of diversity in the dataset results in a biased model. Since phishing websites are often short-lived, collecting them is also a challenge. Consequently, very few phishing webpage dataset repositories exist to date. No single repository comprehensively consolidates all phishing elements corresponding to a phishing webpage, namely, URL, webpage source code, screenshot, and related webpage resources.
   This paper introduces a resource collection tool designed to gather various resources associated with a URL, such as CSS, Javascript, favicons, webpage images, and screenshots. Our tool leverages \texttt{PhishTank} as the primary source for obtaining active phishing URLs. Our tool fetches several additional webpage resources compared to \texttt{PyWebCopy} Python library, which provides webpage content for a given URL. Additionally, we share a sample dataset generated using our tool comprising $4,056$ legitimate and $5,666$ phishing URLs along with their associated resources. We also remark on the top correlated phishing features with their associated class label found in our dataset. Our tool offers a comprehensive resource set that can aid researchers in developing effective phishing detection approaches.
\end{abstract}


\section{Introduction}
\label{sec:Introduction}
The proliferation of mobile devices and the expanding accessibility of Internet services have surged to encompass $66.3\%$ of the global population~\cite{internet_usage}. This widespread connectivity has led individuals to increasingly depend on the Internet for both personal and professional tasks. These activities encompass a wide range of online engagements, including visiting social media platforms, online shopping, managing bank transactions, and accessing educational resources.

These websites collect various forms of sensitive user information to enhance user experiences. Depending on the nature of the platform, this sensitive data may consist of email addresses and passwords for e-commerce and social media sites or more detailed account and debit/credit card information for online banking~\cite{phishing_sharing}. Safeguarding this sensitive data is paramount, as it must remain private and secure to prevent unauthorized access.

The growing reliance on the Internet also heightens the vulnerability of users' sensitive information to cybercrimes. These malicious activities encompass identity theft, ransomware attacks, data breaches, financial losses, and phishing attacks. Perpetrators employ various techniques such as malware, adware, trojans, botnets, SQL injection, man-in-the-middle attacks, and denial of service to carry out these cybercrimes. Such activities result in substantial economic losses and the theft of individuals' identities~\cite{itgovernanceBiggestPhishing}.

\begin{figure}[t]
    \centering
    \includegraphics[width=0.49\textwidth]{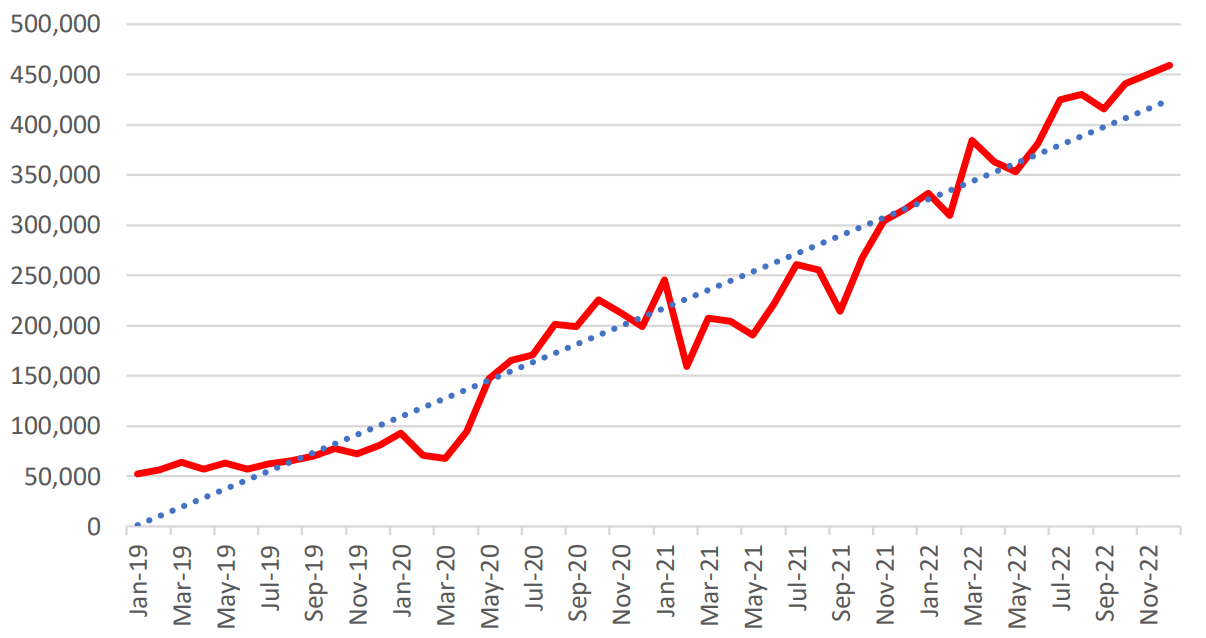}
    \caption{\centering Phishing Attacks from Jan $2019$ to Dec $2022$: APWG~\cite{APWG_REPORT_4_2022}}
    \label{fig:APWG_Report_4_Years}
\end{figure}

Phishing, a deceptive method that combines social engineering and technical subterfuge to steal personal and financial data, has experienced rapid growth, with an annual increase exceeding $150\%$ from $2019$ to $2022$~\cite{APWG_REPORT_4_2022}. Figure~\ref{fig:APWG_Report_4_Years} illustrates the global surge in phishing attacks between January $2019$ and December $2022$, indicating that attackers frequently target reputed brands to create fraudulent webpages.

\begin{figure*}[th]
    \centering
    \includegraphics[width=1.0\textwidth]{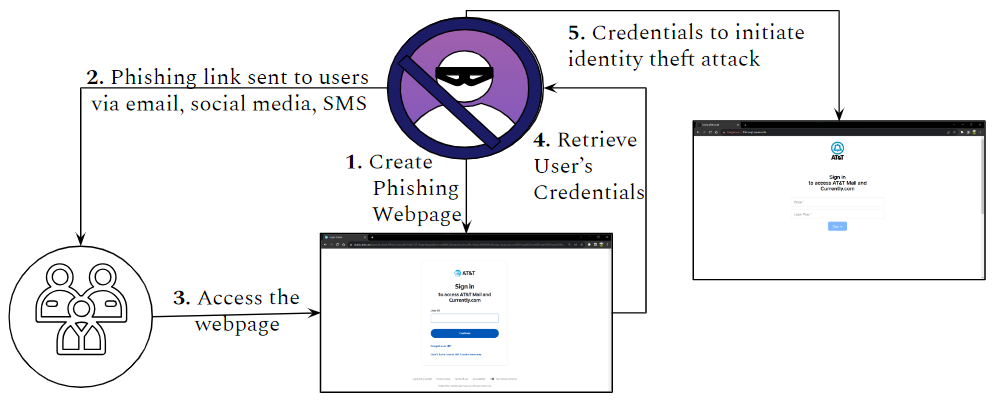}
    \caption{\centering Phishing Attack Lifecycle}
    \label{fig:Phishing_Attack_Lifecycle}
\end{figure*}

Figure~\ref{fig:Phishing_Attack_Lifecycle} outlines the typical stages of a phishing attack, beginning with the attacker targeting a legitimate webpage, crafting a counterfeit phishing page, and disseminating the phishing link via SMS, social media, or email. Unsuspecting users are lured to click on these links, sharing their sensitive information, which is then redirected to the attacker. The attacker may subsequently use these stolen credentials for identity theft or demand a ransom.

Cybersecurity researchers strive to protect users' sensitive information by detecting zero-day phishing attacks. To achieve this, various methods have been developed, including traditional list-based approaches~\cite{cao2008anti}, heuristic-based approaches~\cite{ma2009beyond}, webpage content similarity-based techniques~\cite{zhang2007cantina}, webpage screenshot similarity-based methods~\cite{afroz2011phishzoo}, and machine learning (ML)~\cite{sharma2019phishalert, jain2019machine} and deep learning (DL)~\cite{le2018urlnet, abdelnabi2020visualphishnet} approaches.

List-based approaches involve maintaining databases containing legitimate and phishing URLs and classifying input URLs based on matches with these databases. However, they are ineffective at detecting zero-day phishing URLs~\cite{cao2008anti}. Heuristic-based methods~\cite{ma2009beyond} create pattern heuristics based on observations of phishing and legitimate URLs and webpage content, which vary depending on the dataset.

Visual elements such as webpage screenshots~\cite{afroz2011phishzoo} and logos~\cite{lin2021phishpedia} also contribute to phishing detection by identifying visual patterns within webpages. ML-based approaches~\cite{sharma2019phishalert, jain2019machine} leverage various classifiers trained on datasets of URLs and webpage content to extract relevant features and construct \textit{rule sets} or \textit{decision trees}~\cite{freitas2014comprehensible}. DL-based methods~\cite{le2018urlnet}, on the other hand, automatically extract features without requiring predefined sets, classifying new URLs as legitimate or phishing based on their characteristics.

The existing approaches encompass diverse dataset samples, including URL-based for both phishing and legitimate URLs, content-based approaches utilizing both URL and webpage content, and visual-based methods employing logos, webpage layout, and URL domain to analyze features to efficiently classify a zero-day phishing URL. Nonetheless, the existing repositories~\cite{tranco_db, openphish, millersmiles} fail to offer a comprehensive set of webpage resources for the required samples of phishing and legitimate webpages. Therefore, to tackle the challenges associated with collecting datasets comprising a wide set of resources, we propose a tool to provide relevant resources associated with a landing webpage for any given URL.

The remainder of the paper is organized as follows. Section~\ref{sec:Related_Work} describes the existing open-source repositories researchers use to collect legitimate and phishing samples (URLs, webpage content, and webpage screenshots). Also, we contrast with an existing library~\cite{pywebcopyGitHub} that downloads the webpage content for a given URL. Section~\ref{sec:Proposed_Work} discusses the flowchart of our proposed tool, while Section~\ref{sec:Experiment_Setup_Results_And_Analysis} describes the set-up of our tool along with the observations and analysis. Finally, we conclude with some future scope in Section~\ref{sec:Conclusion}.



\section{Related Work and Our Contributions}
\label{sec:Related_Work}
Many publicly available repositories provide phishing and legitimate samples (URLs, webpage content, and webpage screenshots) as detailed in Table~\ref{tab:Dataset_Repositories}. The currently available open-source repositories may contain samples with a less diverse range of phishing features, which can eventually affect the performance of ML classifiers and DL algorithms in detecting \textit{zero-day} phishing URLs. These open-source repositories play a pivotal role in aiding researchers in expanding their dataset collections with samples containing diverse and informative features.

\begin{table}[t]
\centering
\caption{Dataset Repositories}
\label{tab:Dataset_Repositories}
\begin{tabular}{clp{0.15cm}p{0.15cm}p{0.15cm}p{0.15cm}p{0.15cm}p{0.15cm}p{0.15cm}p{0.15cm}}
\toprule
 &  & \multicolumn{8}{c}{\textbf{Repositories}} \\
\cmidrule{3-10}
& & \rotatebox{90}{\textit{Alexa}~\cite{alexa_db}} & \rotatebox{90}{\textit{Tranco}~\cite{tranco_db}} & \rotatebox{90}{\textit{Common Crawl}~\cite{common_crawl}} & \rotatebox{90}{\textit{PhishTank}~\cite{phishtank}} & \rotatebox{90}{\textit{UCI}~\cite{UCI1}} & \rotatebox{90}{\textit{Mendeley}~\cite{mendeley}} & \rotatebox{90}{\textit{OpenPhish}~\cite{openphish}} & \rotatebox{90}{\textit{Millersmiles}~\cite{millersmiles}} \\ \cmidrule{1-10}
\multirow{2}{*}{\textbf{\makecell{Sample\\Class}}} & \textit{Legitimate} & \ding{51} & \ding{51} & \ding{51} & \ding{51} & \ding{51} & \ding{51} & & \\
& \textit{Phishing} & & & & \ding{51} & \ding{51} & \ding{51} & \ding{51} & \ding{51} \\ \cmidrule(r){1-2} \cmidrule(l){3-10}
\multirow{3}{*}{\textbf{\makecell{Sample\\Type}}} & \textit{URLs} & \ding{51} & \ding{51} & \ding{51} & \ding{51} & \ding{51} & \ding{51} & \ding{51} & \ding{51} \\
& \textit{Webpage Content} & & & \ding{51} & & \ding{51} & \ding{51} & & \\
& \textit{Webpage Screenshot} & & & & \ding{51} & & & & \\
\bottomrule
\end{tabular}
\end{table}

A well-known repository for collecting phishing URLs is PhishTank~\cite{phishtank} due to its consistent updates, including labelling of zero-day phishing URLs by reputed cybersecurity analysts. Initially, Alexa~\cite{alexa_db} held the top position as a repository for collecting the top one million legitimate domains. However, since it ceased its services, Tranco~\cite{tranco_db} became an alternative source for obtaining the same data. Besides these renowned repositories, OpenPhish~\cite{openphish}, Mendeley~\cite{mendeley}, and Millersmiles~\cite{millersmiles} provide zero-day URLs marked as phishing, with Mendeley~\cite{mendeley} even offering the webpage content for each of the URLs. For obtaining legitimate URLs and webpage contents, Commoncrawl~\cite{common_crawl} is also a preferred choice.

Compiling these samples articulates the approach researchers adopt for detecting zero-day phishing attacks. The approaches involve extracting features from the URL, webpage content, webpage screenshots, or combinations thereof, contingent upon whether the phishing detection solution leans toward a content-based or visual-based approach.

To our knowledge, a library named \texttt{pywebcopy} takes a URL as input and retrieves the available resources, subsequently saving them to a local directory. However, this library encounters numerous unresolved issues cataloged on its GitHub repository~\cite{pywebcopyGitHub}. Table~\ref{tab:comparisonWithPywebcopy} delineates the specific challenges faced by \texttt{pywebcopy}, contrasting them with those effectively addressed by our tool. A few distinguishing features distinguishing our tool with the existing library \texttt{PyWebCopy} are as follows:

\begin{enumerate}
    \item The folder structure is not well-organized in \texttt{PyWebCopy}, making it difficult to locate the landing webpage file during processing for webpage content. In contrast, our tool stores each resource in its respective directories, making locating the landing webpage source code easy.
    \item \texttt{PyWebCopy} downloads only the source code of the landing webpage for a given URL but our tool provides all the resources for the webpage.
    \item Also, for many webpages containing Captcha, \texttt{PyWebCopy} could not get permission to scrape the webpage, resulting in errors like \texttt{Forbidden}, \texttt{Not Found}, and if the webpage gets downloaded, then it contains an encoded content. However, this is not the case in our tool.
    \item In terms of the time complexity, \texttt{PyWebCopy} takes $1537.756$ seconds whereas our tool takes $5054.288$ seconds to download the resources for $100$ URLs. This is due to the fact that our tool parses through the webpage to locate, request, and download the associated resources.
\end{enumerate}

\begin{table}[t]
    \centering
    \caption{Contrasting with Existing Library}
    \label{tab:comparisonWithPywebcopy}
    \begin{tabular}{C{1.8cm} C{0.2cm} C{0.2cm} C{0.2cm} C{0.2cm} C{2.2cm}}
    \toprule
        \multirow{10}{*}{\textbf{Tools}} & \multicolumn{5}{c}{\textbf{Contrasting Attributes}} \\ \cmidrule(l){2-6}
        & \rotatebox{90}{File Not Found} & \rotatebox{90}{Content Forbidden} & \rotatebox{90}{Favicon Download} & \rotatebox{90}{Screenshot Download} & \rotatebox{90}{File Structure} \\ \midrule
        PyWebCopy~\cite{pywebcopyGitHub} & \ding{51} & \ding{51} &  &  & Unstructured \\
        Our Tool~\cite{developedToolGitHub} &  & & \ding{51} & \ding{51} & Structured \\ \bottomrule
    \end{tabular}
\end{table}

\emph{Our Contributions}: In this paper, we make two significant contributions to the field of phishing detection and resource collection:
\begin{enumerate}
    \item We introduce a webpage resource collection tool, designed for gathering a wide range of resources associated with a URL, including CSS, Javascript, favicons, webpage images, and webpage screenshots. This tool is a comprehensive solution to aid researchers in developing effective phishing detection approaches.
    \item We provide a dataset consisting of $4,056$ legitimate and $5,666$ phishing webpages and associated resources~\cite{developedToolGitHub} 
\end{enumerate}

\section{Proposed Dataset Collector Tool}
\label{sec:Proposed_Work}

\begin{figure*}[thp]
    \centering
    \includegraphics[width=1.0\textwidth]{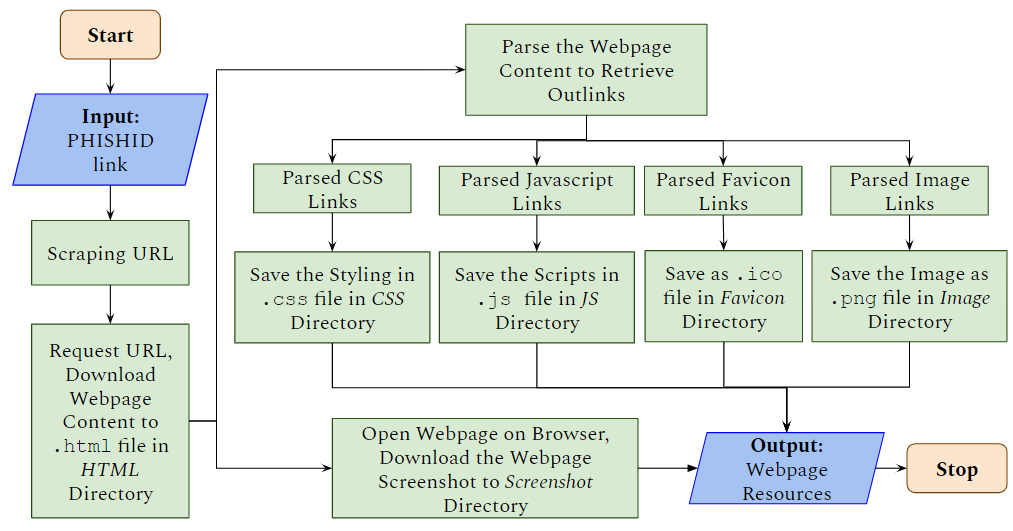}
    \caption{Webpage Resource Collection}
    \label{fig:Dataset_Collection_Flowchart}
\end{figure*}

Existing open-source phishing webpage repositories serve as pivotal resources for collecting lists of zero-day phishing URLs, with some of these repositories even offering webpage content~\cite{common_crawl} and webpage screenshots~\cite{phishtank}. Nevertheless, these repositories only provide some of the resources researchers utilize to identify zero-day phishing URLs. Attackers often employ various methods to bypass existing phishing detection tools to pursue their goals. Therefore, researchers also analyze various webpage elements like CSS, JavaScript, favicons, and logos to identify distinctive features present within the webpages. For instance, in the literature, we observe that phishing webpages replace a portion of legitimate webpage content with an exact screenshot obtained through the CSS \texttt{background-image} and \texttt{z-index} properties. Additionally, attackers may replace logos and favicons with look-alike images by updating the links in the HTML code. Researchers are addressing these emerging tactics attackers employ through ML and DL approaches.

To facilitate phishing detection research, we propose a tool that automates downloading of various related resources associated with a given URL. These resources would include the URL itself, HTML webpage content, styling elements, scripts, favicon images, and webpage screenshots, allowing researchers to conveniently choose the necessary resources for their research approaches. Figure~\ref{fig:Dataset_Collection_Flowchart} outlines the sequential stages through which our tool progresses in downloading the webpage resources. In our study, we leverage PhishTank~\cite{phishtank} as it is an open-source repository widely adopted in phishing webpage detection research. Additionally, the PhishTank repository comprises phishing and legitimate URLs.

From PhishTank's interface, we can generate PHISHID list for corresponding URLs. Scraping of the website thus starts from the PHISHID\footnote{\label{phishid_footnote}\url{https://phishtank.org/phish\_detail.php?phish\_id=PHISHID}} .

Firstly, the algorithm scrapes the PHISHID link\footref{phishid_footnote} containing the URL, the URL is scraped and is requested using various Python libraries to obtain the landing webpage source code. The tool creates a directory named PHISHID on the local system. This directory contains a set of subdirectories, specifically: \textit{CSS, Favicon, HTML, Images, Javascript}, and \textit{Screenshots}. Each subdirectory contains respective files belonging to the landing webpage. For instance, the HTML directory stores the landing webpage source code with the \texttt{.html} extension. Meanwhile, the remaining subdirectories, such as \textit{CSS, Favicon, Images, Javascript}, and \textit{Screenshots}, are populated by downloading the respective resources linked within the landing webpage content. To illustrate further, the Favicon directory contains a \texttt{.ico} file containing the favicon image of the landing webpage. The landing webpage source code contains various tags that are specific to each of the resource outlinks which are parsed using \texttt{BeautifulSoup} Python library~\cite{beautifulsoupRef}. The tool then initiates requests to these parsed links and downloads each of them to their respective directories. Furthermore, the webpage screenshot of the landing webpage is obtained by capturing a snapshot of the webpage visible on the browser and stored in \textit{Screenshot} directory. Finally, the resulting directory comprises comprehensive folders and files for the accessed resources, including the URL, webpage content, screenshots, favicons, CSS, and Javascript files. These resources prove necessary for phishing detection approaches based on URL, content, or visual scrutiny.

\section{Tool Setup, Observations and Analysis}
\label{sec:Experiment_Setup_Results_And_Analysis}
In this section, we discuss the tool setup, and the dataset collected using the tool with observed insights followed by an analysis of the features.
\subsection{Tool Setup}
\label{sec:Experimental_Setup}
Our tool is developed in the Python programming language. It contains a set of libraries that help in requesting the URLs, parsing the HTML webpages to obtain necessary tags, and downloading the favicons and screenshots. This research focused on PhishTank as the primary source for collecting legitimate and live phishing URLs. To execute the automated process for URL collection, we harnessed the \texttt{selenium} framework that systematically navigates through each PhishTank webpage link\footref{phishid_footnote} with a specific PHISHID and scrapes the URL using XPath expressions. The scraped URLs are then recorded in a spreadsheet for further analysis.

Each of the scraped URLs is requested using Python \texttt{requests} library and the corresponding HTML webpage code is downloaded as \texttt{.html} extension to a local directory named \textit{HTML}. The downloaded HTML code is then parsed, using \texttt{BeautifulSoup} library, to acquire the relevant resources like favicon, CSS, and Javascript. The requests to these webpage resources are initiated using libraries such as \texttt{urllib, urlparse,} and \texttt{PIL}. These webpage resources are collected as follows:

\begin{description}
    \item[Javascript] In the downloaded HTML code, the Javascript is either present within the HTML webpage called the \textit{inline Javascript}, or is linked using the \texttt{<script>} tag to an \textit{external Javascript}. These \texttt{<script>} tags are present in the \texttt{<head>} and \texttt{<body>} sections of the HTML code. The inline Javascript is parsed using \texttt{BeautifulSoup.find\_all(`script')} method in the downloaded \texttt{.html} file and the script is written to the \texttt{.js} file in the \textit{Javascript} directory. Similarly, the external Javascript content is parsed using \texttt{BeautifulSoup.find\_all(`script')} by conditioning it with \texttt{src} that returns a \texttt{.js} link. The obtained link to the external Javascript file is requested and the content is written to a \texttt{.js} file present in \textit{Javascript} directory.

    \item[CSS] HTML webpages contain stylings on the HTML tags to describe and present the layout of a webpage on a web browser. There are three types of CSS in an HTML webpage namely, \textit{inline, internal} and \textit{external}. The inline CSS uses the \texttt{style} attribute inside HTML tags and specifies the styles for the respective tags. The internal CSS, \texttt{<style>} tag in the \texttt{<head>} section specifies the tag stylings inside the HTML document itself. The external CSS is linked using the \texttt{<link>} tag that contains the source of the external CSS file. Our tool parses the \texttt{style} attribute in the HTML tags and writes it to the \texttt{.css} file present in the \textit{CSS} directory, also the internal CSS is parsed using \texttt{BeautifulSoup.find\_all(`style')} method and the obtained styles are appended to the \texttt{.css} file. The external CSS is also located using the \texttt{BeautifulSoup.find\_all(`link')} method by conditioning it with the \texttt{rel, href} attributes and the link to the external \texttt{.css} file is obtained and requested to obtain the styles which are then appended to the \texttt{.css} file.

    \item[Favicon] The HTML code is parsed to obtain the link to the favicon for the webpage using links connected to defining the icons in a webpage like: \texttt{icon}, \texttt{apple-touch-icon}, \texttt{shortcut icon}, \texttt{mask icon}, \texttt{fluid icon}, \texttt{manifest}, and \texttt{yandex-tableau-widget}. The algorithm then requests the favicon URL and saves it as a \texttt{.ico} file in the \textit{Favicon} directory. 

    \item[Webpage Screenshot] To obtain the screenshot, we load the HTML document of the landing webpage on a web browser using \texttt{webbrowser} library. The algorithm uses the \texttt{PIL} library to capture the visible portion of the webpage as a screenshot and stores it in a \textit{Screenshot} directory. The captured screenshot is in the form of a binary \texttt{.PNG} image and is converted into a \texttt{PIL} Image using the \texttt{Image.open()} method from the \texttt{PIL} library.

    \item[Images] The tool parses the HTML code to locate the available images in the webpage using \texttt{BeautifulSoup.find\_all(`img')} method. The algorithm iterates through each \texttt{<img>} tag found in the HTML code. The \texttt{<img>} tag contains an attribute named \texttt{src} that contains the link to the image file, and it is obtained using \texttt{img\_tag.get(`src')}. The code checks whether the imgSource URL starts with \texttt{http} to classify it as an absolute URL (if it starts with \texttt{https}) or relative (if it doesn't start with \texttt{https}). If it's not an absolute URL, the relative path is concatenated with the base URL using \texttt{urljoin()} and we obtain the entire URL path to the image file. The obtained link is then requested, and the image is downloaded to the \textit{Image} directory.
\end{description}

\begin{figure}
    \centering
    \includegraphics[width=0.49\textwidth]{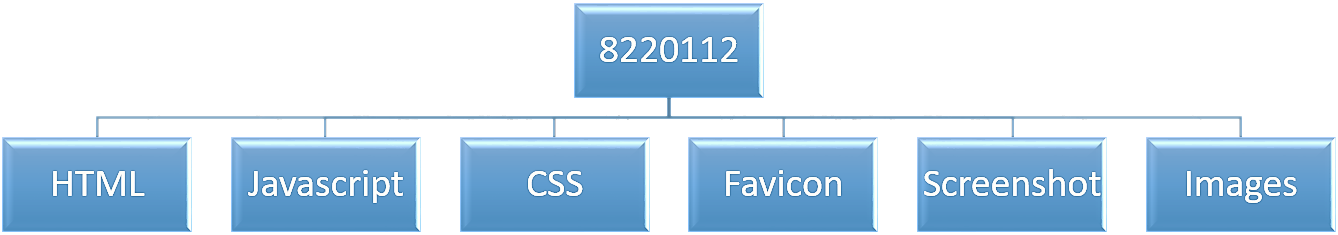}
    \caption{\centering Directory Structure of the Downloaded Webpage with PHISHID: $8220112$}
    \label{fig:8220112_file_structure}
\end{figure}

Let's examine the above process of collecting the webpage resources with an example. A webpage with URL \texttt{http://paypal-net.com/} is obtained by replacing PHISHID in the link\footref{phishid_footnote} with $8220112$. The tool requests the URL and downloads the landing webpage source code. Using \texttt{BeautifulSoup} library, the tool parses and downloads resources like CSS, Javascript, Favicon and Images to their respective directories. Using \texttt{webbrowser} and \texttt{PIL, Image} libraries, we open and download the landing webpage screenshot that is visible on the browser. The file structure of the scraped URL with PHISHID $8220112$ is represented in Figure~\ref{fig:8220112_file_structure}.

\subsection{Observations}
\label{sec:Experiment_Results}

\begin{figure}[b] 
    \centering
    \includegraphics[width=0.49\textwidth]{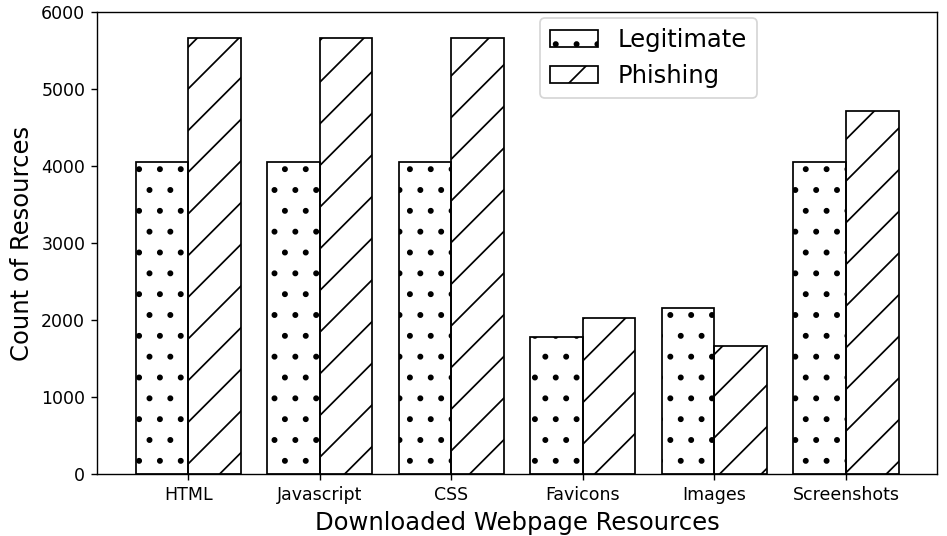} 
    \caption{\centering Statistics of Resources in the Collected Dataset}
    \label{fig:Collected_Dataset}
\end{figure}

In our study, we gathered a dataset encompassing phishing and legitimate URLs from PhishTank~\cite{phishtank}, employing web automation and web scraping techniques. The dataset collection was initiated on July $9$, $2023$ and continued until August $10$, $2023$. This dataset serves as a fundamental component of our research, enabling us to conduct a thorough and in-depth analysis within the scope of our study. Figure~\ref{fig:Collected_Dataset} depicts a bar plot in which the $Y$-axis represents the total count of the webpage resources for both phishing and legitimate URLs and the $X$-axis represents the count of the total number of such resources downloaded for each of the entire dataset for $4,056$ and $5,666$ legitimate and phishing URLs, respectively.



\begin{figure*}[t]
    \centering
    \includegraphics[width=1.0\textwidth]{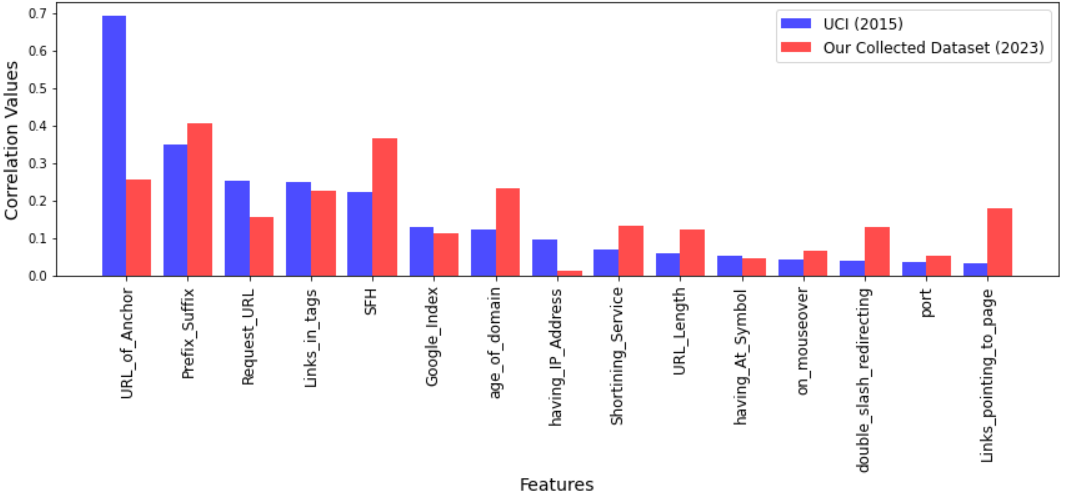}
    \caption{Feature Correlation with the Class Label (Phishing or Legitimate) for UCI and Our Collected Dataset}
    \label{fig:correlation_between_UCI_and_Our_dataset}
\end{figure*}

The percentage of favicons downloaded for the phishing URLs is notably less in comparison to the total count of the phishing webpages. This is due to the fact that phishing webpages are created for gathering sensitive information without much effort applied to create new favicons for a phishing webpage. On the other hand, trustworthy websites usually invest in their branding, which includes creating distinctive and recognized favicons. Additionally, attackers usually apply less effort to create phishing webpages by not spending huge amounts for designing logos and favicons on their target webpages~\cite{das2022modeling}. The total number of webpage screenshots downloaded for the phishing webpages is less because a few webpages are marked as \textit{Dangerous} by the webbrowser and thus, these screenshots are not stored in the respective directories.

Compared with the existing library \texttt{PyWebCopy}, our tool provides a structured directory structure for storing all the relevant resources to the respective subdirectories. Moreover, the directory structure in the \texttt{PyWebCopy} for a given URL contains multiple levels of subdirectories, making it difficult to locate the exact location of the downloaded webpage.



\subsection{Analysis}
\label{sec:Experimental_Analysis_with_Feature_Correlation}

In this section, we delve into an analysis of the correlation of features with class labels (phishing and legitimate), comparing the UCI dataset~\cite{UCI1} and our collected dataset. The UCI dataset consists of $11,055$ instances with $30$ features extracted based on URL, webpage content and third-party components. We extracted the same set of features for our collected dataset. Figure~\ref{fig:correlation_between_UCI_and_Our_dataset} illustrates the correlation values of $16$ most correlated features with the class labels (phishing or legitimate). 

According to the observation, Figure~\ref{fig:correlation_between_UCI_and_Our_dataset} highlights a few insights based on the correlation of features for the $2015$ UCI dataset and for our collected dataset. The increased correlation of the \textit{URL\_of\_Anchor} feature in the UCI dataset suggests that the \texttt{<a>} tags in the UCI dataset frequently consisted \texttt{href=`\#' or `\#content' or `Javascript::void()'} links. However, in our collected dataset, such anchor tags have become less prevalent, replaced by domain outlinks resembling legitimate URLs.

Similarly, the \textit{age\_of\_domain} feature exhibits a high correlation with the classification of webpages as phishing or legitimate for our collected dataset. This is due to advancements in existing tools, which can now accurately classify live phishing webpages in a shorter time compared to their performance in $2015$ for the UCI dataset. Additionally, the \textit{URL\_Length} feature displays increased correlation with webpage classification in our collected dataset as opposed to $2015$. This shift may be attributed to improved phishing detection methods that can differentiate between original URLs and those using shortening services. In contrast, the UCI dataset's heuristics relied on URL length exceeding $75$ characters for classifying a URL as phishing. The \textit{having\_IP\_Address} feature exhibits a high correlation in the UCI dataset, primarily because, in $2015$, many phishing webpages used IP addresses instead of domain names for hosting. However, as attackers evolved their tactics, phishing webpages began utilizing \texttt{http/https} in recent times and are also hosted on compromised domains. Consequently, analyzing such webpages now demands a more comprehensive examination of other features.

To summarize, our tool offers a comprehensive set of associated resources for a given URL, simplifying the selection process for researchers in developing their approaches. Also, the phishing features evolve with time as attackers improve their tactics to bypass the existing phishing detection tools. The correlation of the features with its class labels drastically changes with the design of phishing webpages over the years. Hence, feature \emph{selection} algorithms need to be employed to get more information and develop a model with high accuracy in correctly classifying a new suspicious URL.





\section{Conclusion and Future Work}
\label{sec:Conclusion}
In this paper, we proposed a resource collection tool that collects various webpage resources such as CSS, Javascript, favicon, webpage images, and screenshots for a given URL. By providing such a diverse webpage resource set, we aim to empower researchers to develop detection approaches to effectively counter the zero-day phishing attacks. Our too outperforms \texttt{PyWebCopy} library in acquiring not only landing webpage content but also the relevant resources associated with the landing webpage. Additionally, we contribute the dataset containing $4,056$ legitimate and $5,666$ phishing webpage resources~\cite{pywebcopyGitHub}. Our tool and dataset will aid cybersecurity researchers as a resource collector for datasets due to its comprehensive inclusion of all the necessary resources linked to the landing webpage of a given URL.

In our present tool, we leveraged PhishTank as the primary source for collecting active phishing URLs. However, in forthcoming versions of the tool, we intend to broaden our sources to include other repositories, thereby enhancing the diversity of our dataset in terms of phishing features. Additionally, our tool captures only the visible portion of the landing webpage as a screenshot. Although there are extensions available that can capture the entire webpage screenshot, we have refrained from using them due to third-party dependency on our tool. Instead, we plan to develop our proprietary algorithm to perform this task independently.

\bibliographystyle{IEEEtran}
\balance
\bibliography{references}

\end{document}